\documentstyle[]{laa} 

\def\lsim {\ifmmode {\buildrel<\over\sim}                              
	\else {\lower.6ex\hbox{$\buildrel<\over\sim$}}\fi}  
\def\gsim {\ifmmode {\buildrel>\over\sim}                              
	\else {\lower.6ex\hbox{$\buildrel>\over\sim$}}\fi}
 
\begin{document} 
  
  \thesaurus{09     
       (09.03.1; %
	09.09.1; %
	09.10.1; %
	08.06.2; %
	08.13.2) %
       } 
  \title{Jet driven molecular outflows in Orion} 
  
  
  \author{A. Rodr\'{\i}guez$-$Franco$^{1,2}$ and J. Mart\'{\i}n$-$Pintado$^2$ and 
	  T.L. Wilson$^{3,4}$ 
     } 
 
  \offprints{A. Rodr\'{\i}guez$-$Franco to the IGN address} 
  
  \institute{$^1$Departamento de Matem\'atica Aplicada$\,$II, Secci\'on  
       departamental de Optica, Escuela Universitaria de Optica,  
       Universidad Complutense de Madrid. Av. Arcos de Jal\'on s/n.  
       E-28037 Madrid. Spain\\ 
       $^2$Observatorio Astron\'omico Nacional (IGN), 
       Campus Universitario, Apdo. 1143, E-28800,  
       Alcal\'a de Henares, Spain\\
       $^3$Max Planck Institut f\"ur Radioastronomie, Postfach 2024, D-53010 Bonn,
       Germany\\
       $^4$Sub-mm Telescope Observatory, Steward Observatory, The University of Arizona,
       Tucson, AZ, 85721 
       }  
 
  \date{Received, 1999; accepted, 1999} 
  
  \maketitle 
  
  \begin{abstract} 
%
We present high sensitivity and high angular resolution images of the high velocity
($v_{LSR}>30\,$km$\,$s$^{-1}$) CO emission in the
$J=1\rightarrow0$ and $J=2\rightarrow1$ lines of the Orion~KL region. These results reveal
the morphology of the high-velocity CO emission at the most extreme
velocities. High  
velocity emission have been only detected in two regions: BN/KL (IRc2/I) and  
Orion--S.

The Orion--S region contains a very young (dynamical age of $\sim10^3\,$years), very fast 
($\sim110\,$km$\,$s$^{-1}$) and very compact (\lsim$0.16\,$pc) bipolar outflow. From the 
morphology of
the high-velocity gas we estimate that the position of the powering source must be
$20''$ north of FIR$\,4$. So far, the exciting source of this outflow has not been
detected. For the IRc2/I molecular outflow the morphology of the moderate velocity
(\lsim60$\,$km$\,$s$^{-1}$) gas shows a weak bipolarity around
IRc2/I. The gas at the most extreme velocities does not show any  bipolarity around
IRc2/I, if any, it is found $\sim30''$ north from these sources. The blue and redshifted
gas at moderate velocities shows similar spatial distribution with a systematic trend for
the size of the high-velocity gas to decrease as the terminal radial velocity increases.
The same trend is also found for the jet driven molecular outflows L$\,$1448 and 
IRAS$\,03282+3035$. The size-velocity relationship is fitted with a
simple velocity law which considers a highly collimated jet and entrained material
outside the jet moving
in the radial direction.
We also find that most of the CO outflowing at moderate velocities is located at the 
head of the jet. Our results and the spatial distribution and kinematics 
of
the shock tracers in this outflow can be explained if the IRc2/I outflow is
driven by a precessing jet oriented along the line of sight. The implication of these
findings in the evolution of molecular outflows is discussed.

  
  \end{abstract} 
  
%
  
\section{Introduction} 
  
The Orion Molecular Cloud (hereafter OMC$\,$1) has played a central role in the 
study of the formation and evolution of high$-$mass 
stars. This molecular cloud harbors massive stars in the stage of energetic mass loss. Two
regions contain high-velocity (HV hereafter) molecular outflows, the IRc2/I outflow in
the BN/KL nebula and the Orion--S region. 

The first evidence of massive star formation in the Orion--S region, located
$\sim100''$ south of IRc2 came from the presence of warm gas (Ziurys et al., 1981)
and large dust column densities (Keene, Hildebrand \& Whitcomb, 1982).
Subsequent observations of CO with higher angular resolution showed the presence of a 
long 
filament at moderate velocities ($\leq 30\,$km$\,$s$^{-1}$) which has been interpreted as
a redshifted monopolar jet powered by FIR$\,$4 (Schmid$-$Burgk et al., 1990). H$_2$O
masers have been detected in the vicinity of FIR$\,$4 but they are not related
with the monopolar jet (Gaume et al., 1998). In fact, the H$_2$O masers seem to be 
associated with the very fast 
bullets ($\mid v_{LSR}\mid>50\,$km$\,$s$^{-1}$) found in the compact bipolar 
outflow recently detected by Rodr\'{\i}guez--Franco et al. (1999) which is perpendicular
to the low velocity jet.

Due to its nearness and intense CO emission, the prominent high-velocity CO outflow 
in the KL nebula has been the subject of intense study in several rotational  
transitions of this molecule (Kwan \& Scoville, 1976; Zuckerman, Kuiper \&  
Rodr\'{\i}guez$-$Kuiper, 1976; Wannier \& Phillips, 1977; Phillips et al., 1977; 
Goldsmith et al., 1981; Van Vliet et al., 1981). The first observations showed the
presence of outflowing gas at velocities up to 
$100\,$km$\,$s$^{-1}$ from the cloud velocity. Strong shocks produced by the
interaction of the outflow from the powering source with the ambient material gives rise
to strong emission of vibrationally excited molecular hydrogen H$_2^*$ (see e.g.
Nadeau \& Geballe, 1979), H$_2$O
masers around the source IRc2/I (Gaume et al, 1998) and molecules like SiO, SO, SO$_2$ 
(Wright et al., 1996) produced by shock chemistry. 

In spite of the large numbers of observations of this molecular outflow, its nature is so
far unclear. Observations with high angular resolution of the HV CO emission at moderate
velocities (Erickson et al., 1982; Chernin \& Wright, 1996) have shown a weak bipolarity
of the outflow. Based on these observations, it was proposed that this outflow is a
conical outflow with a wide ($130^{\rm o}$) opening angle, oriented in the 
southwest$\leftrightarrow$northeast direction. This is powered by source I (Chernin \& 
Wright,
1996). However the wide opening angle model cannot account for a number of observational 
facts such as the spatial distribution and
kinematics of the high and low velocity H$_2$O and the SiO masers
(Greenhill et al., 1998; Doeleman et al., 1999). Furthermore, the morphology of
the HV CO emission at intermediate velocities shows that the outflow is offset from
IRc2/I by $10''$ indicating that a complex channeling of the HV gas is needed if IRc2/I
is the powering source (Wilson et al., 1986). Recently Rodr\'{\i}guez--Franco et al. 
(1999) has mapped the
IRc2/I molecular outflow with high sensitivity. The spatial distribution of the most
extreme velocity shows the lack of bipolarity around IRc2/I as expected from the wide 
opening angle model. These maps also reveal a ring-like structure of high-velocity bullets
which also are difficult to account for if the outflow has a wide opening angle. 
Rodr\'{\i}guez--Franco et al. (1999) have proposed that the ring of the HV CO bullets and 
the distribution and 
kinematics of
the H$_2$O masers could be explained if the outflow is driven by a precessing jet
oriented along the line of sight. If this is the explanation, this will allow us to
specify the main entrainment mechanism in molecular outflows (see e.g. Raga \&
Biro, 1993; Cabrit, 1995).

In this paper we present new maps of the CO emission and analyze in detail the 
morphology of the CO emission at moderate and extreme
velocities in the IRc2/I and Orion--S molecular outflows, and present new arguments
supporting the idea that the Orion outflows are young and driven by high velocity jets.  

%
  
\section{Observations and results} 
 
The observations of the $J=2\rightarrow1$ and the $J=1\rightarrow0$ lines of  
CO where carried out with the IRAM 30$-$m telescope at Pico Veleta (Spain).  
Both rotational transitions were observed simultaneously with SIS  
receivers tuned to single side band (SSB) with an image rejection of  
$\sim 8\,$dB. The SSB noise temperatures of the  
receivers at the rest frequencies for the CO lines were 300$\,$K and 110$\,$K 
for the $J=2\rightarrow1$  
and the $J=1\rightarrow0$ lines, 
respectively. The half power beam width (HPBW) of the telescope was $12''$ for  
the  
$J=2\rightarrow1$ line and $24''$ for the $J=1\rightarrow0$ line. As  
spectrometers, we used two filter banks of $512\times1\,$MHz that provided a  
velocity  
resolution of 1.3 and 2.6$\,$km$\,$s$^{-1}$ for the $J=2\rightarrow1$ and the  
$J=1\rightarrow0$ lines of CO, respectively. The 
observation procedure was position switching with the reference taken at a fixed 
position located 15$'$ away in right ascension. The mapping was carried out by   
combining 5 on$-$source spectra with 
one reference spectrum. The typical integration times were 20$\,$sec for the  
on-positions 
and 45$\,$sec for the reference spectrum. The rms sensitivity of a single   
$J=2\rightarrow1$ spectrum 
is 0.6$\,$K. Pointing was checked frequently on  
nearby 
continuum sources (mainly Jupiter) and the pointing errors were  
\lsim4$''$. The calibration 
was made by measuring sky, hot and cold loads. The line intensity scale  
is in units of main beam brightness temperature, using main beam 
efficiencies of 0.60 for the  
$J=1\rightarrow0$ line and 0.45 for the $J=2\rightarrow1$ line.

\begin{figure*} 
   \vspace{7.5cm}
   \includegraphics{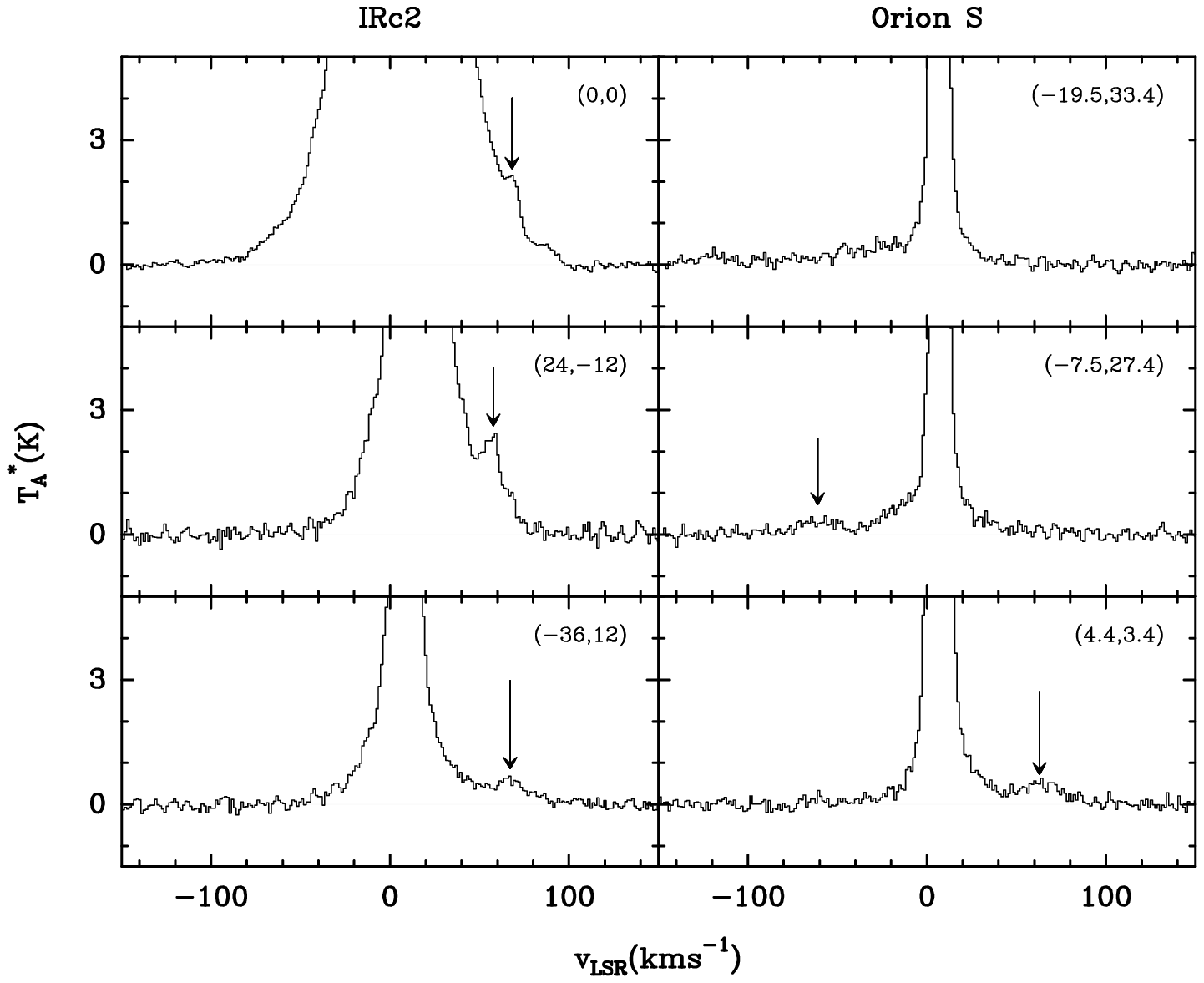}
   \caption{Sample of CO $J=2\rightarrow1$ line profiles taken towards  
	selected positions in the vicinity of the molecular outflows IRc2  
	(left panels) and Orion$-$S (right panels). The vertical arrows show  
	the location of high-velocity ``bullets'' similar  
	to those observed in some bipolar outflows driven by low mass stars. 
	The offsets are relative to the position of IRc2 and FIR$\,$4 for the IRc2  
	and the Orion--S molecular outflows respectively.   
       }  
     \label{fig:espectros-co} 
  \end{figure*}

We have made an unbiased search for high-velocity molecular gas in OMC$\,$1 by  
mapping with 
high sensitivity the $J=1\rightarrow0$ and $J=2\rightarrow1$ 
lines of CO in a region 
of $14'\times14'$ around IRc2/I.  
In this article we will analyze the very high-velocity 
gas associated to molecular outflows. The widespread CO emission with 
moderate velocities 
($\mid v_{LSR}-9\mid$\lsim40$\,$km$\,$s$^{-1}$) will be discussed 
elsewhere (Mart\'{\i}n$-$Pintado \& Rodr\'{\i}guez$-$Franco, 2000).  
In our CO maps we have only detected the known molecular outflows with
$\mid v_{LSR}-9\mid$\gsim30$\,$km$\,$s$^{-1}$, IRc2 and  
Orion$-$S. Figure 
\ref{fig:espectros-co} shows a sample of line profiles taken towards  
selected positions  around IRc2 and Orion$-$S.

\section{High-velocity gas around IRc2/I} 
 
\subsection{The morphology} 
 
\begin{figure*}
   \vspace{7.5cm}
   \includegraphics{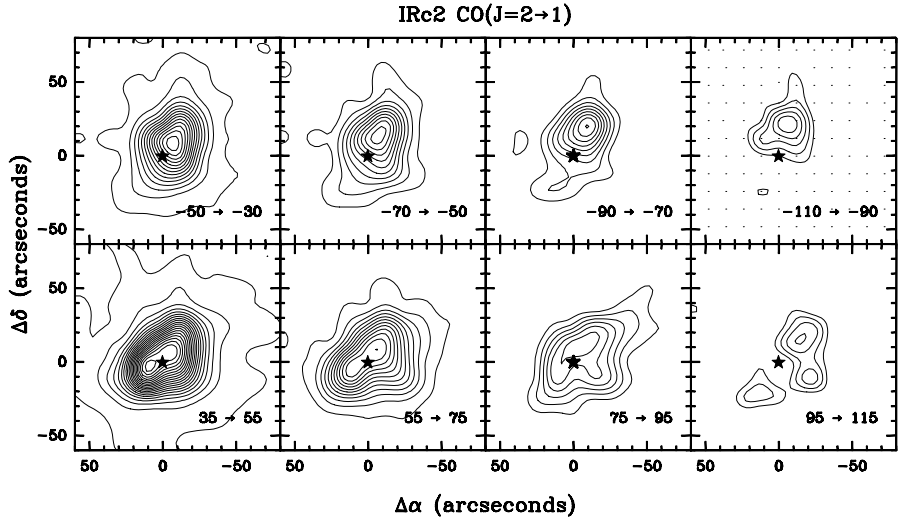}
   \caption{Integrated intensity maps of the CO $J=2\rightarrow1$ line for radial 
	velocity intervals of 20$\,$km$\,$s$^{-1}$ for the blueshifted (upper  
	panels) and redshifted (low panels) wings. The radial velocity 
	intervals appear in the low right corner of every panel.The offsets 
	are relative to the position 
	of IRc2 ($\alpha$(1950)=5$^{{\rm h}}$ 32$^{{\rm m}}$ 47.0$^{{\rm s}}$, 
	$\delta$(1950)=--5$^{{\rm o}}$ 24$'$ 20.6$''$) represented  
	in every panel by the filled star. The positions where the spectra  
	were measured are shown by dots on the top right panel.  
	For all panels, the first contour level is 2$\,$K$\,$km$\,$s$^{-1}$.  
	From the left to the right panels the interval between levels are, 
	respectively, 8$\,$K$\,$km$\,$s$^{-1}$, 4$\,$K$\,$km$\,$s$^{-1}$,  
	2$\,$K$\,$km$\,$s$^{-1}$ and 1$\,$K$\,$km$\,$s$^{-1}$. 
       } 
     \label{fig:IRc2-vel} 
  \end{figure*}

Fig. \ref{fig:IRc2-vel} shows the spatial distribution of the CO  
$J=2\rightarrow1$ integrated intensity emission around IRc2 for radial  
velocity intervals of 20$\,$km$\,$s$^{-1}$ for the blueshifted and  
redshifted gas. The 
spatial distribution of the HV gas with moderate velocities  
(\lsim60$\,$km$\,$s$^{-1}$), hereafter MHV, in our data is similar to that observed by  
Wilson et al. (1986).  
Our new maps have better sensitivity. These reveal for the first time the  
spatial  
distribution of the  
high-velocity gas for the most extreme velocities  
($\mid v_{LSR}\mid >60\,$km$\,$s$^{-1}$), hereafter EHV, of the molecular outflow which  
shows a different 
morphology from that of the MHV gas.

The maximum of the CO emission for the blueshifted gas 
always appears northwest of IRc2/I. The offset between the CO maximum and IRc2  
systematically increases 
from 10$''$ for radial velocities of $-$40$\,$km$\,$s$^{-1}$ up to 25$''$ for  
the  
gas at $-$110$\,$km$\,$s$^{-1}$. The location of the redshifted CO emission  
maxima does 
not show such a clear 
trend. At moderate velocities (up to  
75$\,$km$\,$s$^{-1}$), the  
maximum CO 
emission occurs in a ridge with two peaks of nearly equal intensity located  
southeast 
and northwest of IRc2/I. The northwest peak of the redshifted emission occurs at  
the same 
position as the blueshifted peak for lower radial velocities.  
For the most redshifted gas ($\sim$100$\,$km$\,$s$^{-1}$), the CO emission breaks  
up into several condensations located around IRc2/I. 
For this radial velocity, the most intense CO condensation peaks as for the  
blueshifted gas, 
northwest of IRc2, there is 
weaker emission found 20$''$ west and southeast of IRc2/I.

\begin{figure*}  
   \vspace{10.5cm}
   \includegraphics{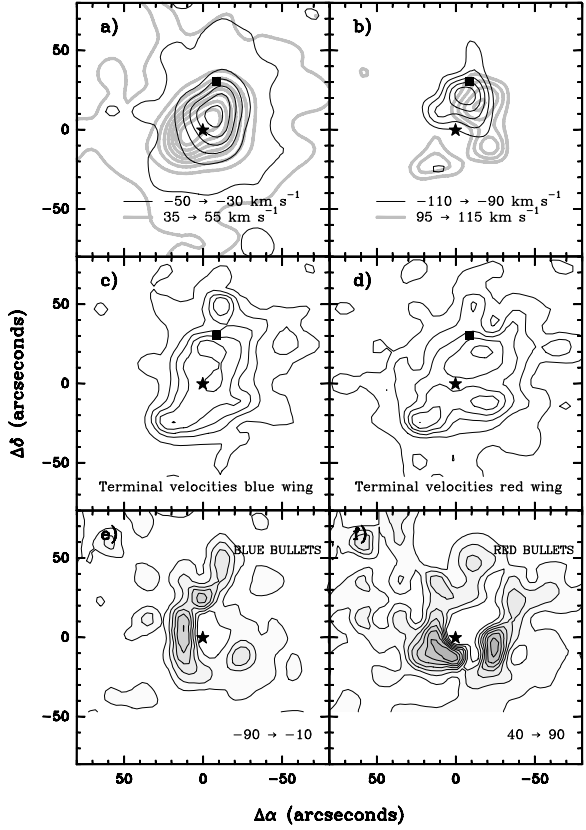} 
	\caption{{\bf a)} Spatial distribution of the integrated emission  
	of the $J=2\rightarrow1$ CO line 
	for blueshifted (solid contours) and redshifted (dashed contours) gas at 
	moderate velocities.  
	The lowest level corresponds to an integrated  
	intensity of 2$\,$K$\,$km$\,$s$^{-1}$; 
	the interval between levels is 20$\,$K$\,$km$\,$s$^{-1}$. 
	\newline            
	{\bf b)} Spatial distribution of the integrated $J=2\rightarrow1$ CO  
	emission for the most extreme velocities of  
	the blueshifted (solid contours) and redshifted (dashed contours) gas.  
	The first  
	level corresponds to  
	2$\,$K$\,$km$\,$s$^{-1}$ and the interval between levels is  
	1$\,$K$\,$km$\,$s$^{-1}$.  
	Note that for the most extreme velocities, the blue and red wing CO  
	maxima appear  
	north of IRc2, which does not support the idea of 
	bipolarity around IRc2/I. 
	\newline 
	{\bf c) \& d)} Spatial distribution of the iso-terminal velocities of  
	the  
	high-velocity  
	gas as measured from the $J=2\rightarrow1$ CO line corresponding to the  
	blue and red wings respectively. For the blueshifted gas, the first    
	level is $-30\,$km$\,$s$^{-1}$ and the interval between levels is  
	$-20\,$km$\,$s$^{-1}$. For the redshifted gas, the first contour level  
	corresponds to 45$\,$km$\,$s$^{-1}$ and the interval to 
	20$\,$km$\,$s$^{-1}$. 
	\newline 
	{\bf e) \& f)} integrated intensity of the CO $J=2\rightarrow1$ line
	emission
	between $-90$ and $-10\,$km$\,$s$^{-1}$, and between 
	$40$ and $90\,$km$\,$s$^{-1}$ for the blue and red HV bullets 
	respectively. The maps have
	been obtained subtracting a Gaussian profile to the broad line wings
	(see Rodr\'{\i}guez--Franco et al., 1999).
	The first contours level is $2\,$K$\,$km$\,$s$^{-1}$, and the interval between
	levels is $7\,$K$\,$km$\,$s$^{-1}$.
	\newline
	For all the panels, the  
	filled star shows the position  
	of IRc2 and the filled square the position of IRc9.} 
     \label{fig:bipolar} 
  \end{figure*}

Fig. \ref{fig:bipolar}a and b summarize  
the distribution 
of the high-velocity molecular gas in Orion for moderate and extreme velocities. 
The new data show that 
the MHV molecular gas around IRc2  
(fig. \ref{fig:bipolar}a) has a very different 
morphology than that of the EHV gas (fig. \ref{fig:bipolar}b). 
At moderate velocities, the CO  
emission shows a very weak bipolarity (if any) around 
IRc2 in the southeast$\leftrightarrow$northwest direction  
(see Fig. \ref{fig:bipolar}a). For radial  
velocities close to 
the terminal velocity, the strongest CO emission (see Fig. \ref{fig:bipolar}b)  
does not show any bipolarity around IRc2/I. The only possible
bipolar morphology is observed around a position 
located $\sim$20$''$ northwest of IRc2 and $10''$ south of IRc9 represented as 
a filled square 
in Fig. \ref{fig:bipolar}.

\subsection{The size-velocity relation in the HV gas}

The terminal velocities of the blue and redshifted MHV gas show similar 
spatial distribution with an elliptical--like shape  
centred in the vicinity of IRc2/I, and a systematic trend of the  
size of HV gas to decrease as a function of the velocity (see Fig. 
\ref{fig:IRc2-vel}). These two 
characteristics are illustrated in Fig. \ref{fig:bipolar}c and d 
where we plot the dependence of the size of the HV CO emission as a  
function of the terminal velocity for  
the red and the blue shifted gas respectively. The contours in these figures  
shows the location of the gas with the same terminal velocity.  
The area enclosed by a given contour level corresponds to  
the region which contains HV molecular gas with terminal velocities  
larger than that of the contour level. 
For moderate velocities (\lsim60$\,$km$\,$s$^{-1}$), the size of HV gas
is $\sim$110$''$ and  
decreases to $\sim$40$''$ for the extreme velocities.  
The HV gas, for most radial velocities, is concentrated in a region 
with an elliptical like
shape centered at ($-4''$, $4''$) with respect to IRc2/I.

\subsection{High-velocity ``Bullets''} 

The detection of these high velocity bullets in the Orion IRc2/I outflow has been 
reported by
Rodr\'{\i}guez--Franco et al. (1999). For completeness, 
we summarize the main characteristics.
Fig. \ref{fig:espectros-co} shows several examples of CO profiles  
towards the IRc2/I region where the location of the HV bullets are shown by vertical 
arrows. 
The typical line-width of the CO HV bullets
is $20-30\,$km$\,$s$^{-1}$, these are distributed in a ring like structure of size
$\sim10''\times50''$ ($0.02\times0.1\,$pc) 
and thickness $12''-20''$ ($0.02-0.04\,$pc) with IRc2 located in the southeast edge of
the HV bullets rings
(see fig. \ref{fig:bipolar}e and f).

\section{High velocity gas around Orion--S} 
 
\subsection{The morphology and characteristics of the high-velocity bipolar  
outflow Orion--S} 
 
\begin{figure*}
   \vspace{7.5cm}
   \includegraphics{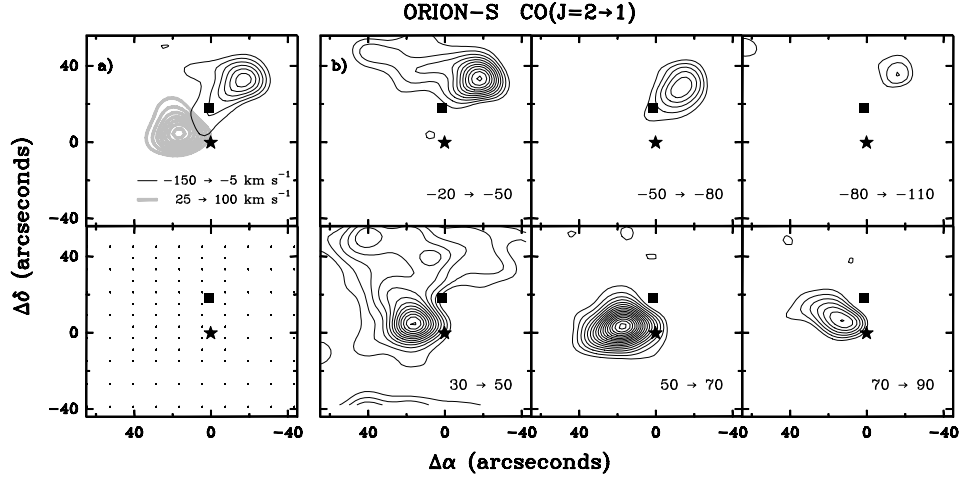}   
   \caption{{\bf a)} The upper panel shows the spatial distribution of the  
	    fast molecular outflow in Orion--S. 
	    The dashed contours show the spatial extension of the  
	    $J=2\rightarrow1$ CO emission for the red wing, and solid 
	    contours represent the blue wing. The first level is 
	    9$\,$K$\,$km$\,$s$^{-1}$ and the interval between levels is 
	    2.5$\,$K$\,$km$\,$s$^{-1}$.  
	    The lower panel shows the positions where the spectra were taken. 
	    The offsets are relative to the FIR$\,$4  
	    ($\alpha(1950)=-5\,^{\rm h} 32\,^{\rm m} 45.9\,^{\rm s}$, 
	    $\delta(1950)=-5^{\rm o} 26' 6''$) represented by a filled  
	    star. The filled square represents the position of the  
	    possible exciting source (see text). 
	    \newline 
	    {\bf b)} Maps of the integrated $J=2\rightarrow1$ CO emission 
	    as a function of the radial velocity in the direction of the 
	    Orion$-$SFBO bipolar outflow for the blue (three upper panels) and  
	    redshifted (three lower panels) wings. 
	    Velocity intervals in $\,$km$\,$s$^{-1}$ are noted in 
	    the lower right corner of every panel. In the velocity intervals  
	    between $-20$ and $-80\,$km$\,$s$^{-1}$ and between 30 and  
	    $70\,$km$\,$s$^{-1}$ the first contour level is  
	    1.4$\,$K$\,$km$\,$s$^{-1}$ and the distance between levels is  
	    0.6$\,$K$\,$km$\,$s$^{-1}$. For the velocity intervals between  
	    $-80$ and $-110\,$km$\,$s$^{-1}$  and between 70 and  
	    90$\,$km$\,$s$^{-1}$ the first contour level is  
	    1$\,$K$\,$km$\,$s$^{-1}$ and the distance between levels is  
	    0.3$\,$K$\,$km$\,$s$^{-1}$.  
       } 
     \label{fig:surarea} 
  \end{figure*}

Moderate high velocity gas ($\leq 30\,$km$\,$s$^{-1}$) have  
been detected in this region in CO and SiO (Schmid--Burgk et al.   
1990; Ziurys \& Friberg, 1987). The CO emission with moderate velocities has 
been associated  
to a monopolar outflow (low velocity redshifted jet) in Orion--S  
discovered by Schmid--Burgk et al. (1990). Rodr\'{\i}guez--Franco et al. (1999)
reported the detection of a new compact and Fast Bipolar Outflow (hereafter Orion-South 
Fast Bipolar Outflow, or Orion-SFBO)
with terminal velocities of $\sim100\,$km$\,$s$^{-1}$.
In this work, we will present the main features of this new outflow. To avoid the
confusion with other HV features in Orion--S, such as H$\,$II region 
(see Mart\'{\i}n--Pintado et al., 2000) and the monopolar outflow (Schmid--Burgk et 
al. 1990) we will consider only the CO emission for radial  
velocities in the range    
$<-5\,$km$\,$s$^{-1}$, and $>25\,$km$\,$s$^{-1}$.   
Fig. \ref{fig:surarea}a shows the integrated intensity map in the  
$J=2\rightarrow1$ rotational transition of CO for the blueshifted (solid  
contours) and the redshifted line wings (dashed contours) of the 
Orion$-$SFBO. This new outflow shows a clear bipolar structure 
in the southeast$\leftrightarrow$northwest direction, just perpendicular to  
the low velocity  redshifted jet.  

The morphology of the Orion$-$SFBO suggests that the axis of the bipolar outflow  
is oriented close to the plane of the sky.  
The blueshifted emission peaks $40''$  
northwest of FIR$\,$4 while redshifted gas has its maximum intensity $15''$  
northeast from that source.
Since the morphology of the bipolar SiO emission (Ziurys \& Friberg, 1987)  
is similar to the CO emission reported in this paper, the SiO emission is very 
likely related to the Orion$-$SFBO rather than to the low velocity monopolar jet 
(Schmid--Burgk et al., 1990).

The structure of the HV gas in the Orion$-$SFBO  
as a function of the radial velocities  
is shown in Fig. \ref{fig:surarea}b. The HV gas in the  
blue and redshifted wings have a different behavior. While the 
redshifted CO emission peak is located at the same spatial location for all radial 
velocities, 
the blueshifted CO peak moves  to larger distance ($\sim10''$, nearly one beam)
from FIR$\,$4  
as the radial velocity increases from $-70$ to $-100\,$km$\,$s$^{-1}$.  
The most extreme velocities in the blue lobe (between $-80$ and 
$-110\,$km$\,$s$^{-1}$) arise from a condensation of  
$\sim11''$ located $36''$ northwest from FIR$\,$4. High velocity bullets (see vertical
arrows in fig. \ref{fig:espectros-co}) also appear in both the blue and the redshifted, 
lobes. A description can be found in Rodr\'{\i}guez--Franco et al. (1999).

\begin{table}[bthp] 
\begin{center} 
\begin{tabular}{ccccc}  
\hline 
     & $v_{mt}$\footnotemark[1] & Mass\footnotemark[2] & Momentum & Energy\\ 
     & (km$\,$s$^{-1}$) &($M_{\odot}$)&($M_{\odot}$$\,$km$\,$s$^{-1}$)& (erg)\\ 
\hline 
\hline 
\multicolumn{5}{c}{Wings}\\ 
\hline 
\hline 
blue & $-140$ & 0.019 & 14.3 & 7.97$\times10^{45}$\\ 
red & 88 & 0.017 & 13.3 & 5.46$\times10^{45}$\\ 
\hline 
\hline 
\multicolumn{5}{c}{Bullets}\\ 
\hline 
\hline 
blue & $-140$ & 3.4$\times10^{-3}$& 0.5 & 6.9$\times10^{44}$\\ 
red & 88 & 8.6$\times10^{-3}$& 0.7 & 5.6$\times10^{44}$\\ 
\hline 
\multicolumn{5}{l}{}\\ \cline{1-2} 
\multicolumn{5}{l}{{\footnotesize $^1$Maximum velocity}}\\ 
\multicolumn{5}{l}{{\footnotesize $^2$For a CO abundance of 
 $10^{-4}$}}\\ 
\multicolumn{5}{l}{{\footnotesize  ~and excitation temperature of 80$\,$K.}} \\ 
\end{tabular} 
\end{center} 
\caption[]{Physical parameter of the molecular outflow Orion--SFBO and the  
	   associated bullets.} 
\label{tab:parasur} 
\end{table}

\subsection{Exciting source}

The source(s) powering the high-velocity gas in Orion--S is, so far, unknown. 
The source FIR$\,$4   
has been proposed to be the exciting source of the low velocity jet 
(Schmid$-$Burgk et al., 1990). However,  FIR$\,$4 cannot be the powering source 
for this outflow, because this 
source is located $20''$ south from the geometrical center defined by
the two lobes (see figure \ref{fig:surarea}a).  
One can use the kinematics and the morphology of the HV gas to estimate the  
position of the exciting source.  
This procedure has been successfully used for outflows associated with low  
mass stars (see Bachiller et al., 1990). From the velocity$-$position diagram 
along the direction of the outflow axis we have estimated  
the location of the possible exciting source by considering that the source  
should be located at the position where the radial velocity changes from blue  
to redshifted. 
The inferred position of the exciting source using this procedure is located $20''$ 
north of FIR$\,$4, and it is 
shown in Fig. \ref{fig:surarea} as a filled square.

In contrast to the outflows powered by low mass star where the exciting source  
appears as strong millimeter emitters and faint centimeter emitters, the source(s) 
of the Orion$-$SFBO has not been detected so far in
the mm or cm continuum emission 
(see e.g. Gaume et al., 1998). The limit of the cm radio continuum emission is 
a factor of 10 smaller than the predicted by the relationship of Anglada et al. 
(1998) for  collisional ionization. This could be due to an underestimation of 
the dynamical age of the outflow or more likely to a time variable jet (see 
section \ref{sec:jet}).

\section{The physical conditions of the HV gas in the IRc2/I outflow and in 
Orion$-$SFBO} 
\label{sec:conditions}

\begin{figure*} 
   \vspace{7.5cm}
   \includegraphics{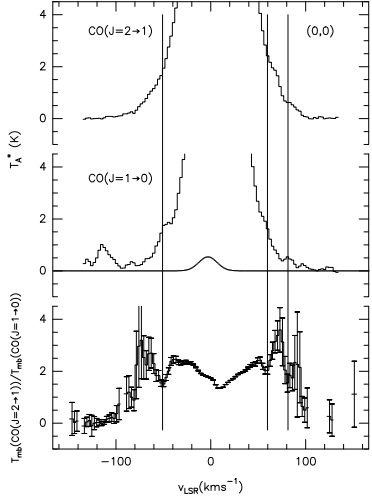} 
   \caption{The lower panel is a plot of the ratio between the intensity of the CO  
            $J=2\rightarrow1$ and $J=1\rightarrow0$ lines. These are shown in the upper 
            and the middle panels, respectively. When the emission is optically thin in both
            lines this is the opacity ratio. Vertical lines are traced for velocities
            where the ratios are almost unity. In these positions an increase of the 
            relative intensity of one of the line profiles is shown. The middle panel 
            also shows the recombination line emission H38$\alpha$ (dotted profile). 
	   } 
     \label{fig:ratio} 
  \end{figure*}

The H$_2$  densities of the HV gas in the IRc2/I outflow are high enough 
($\sim10^5\,$cm$^{-3}$, see Boreiko et al., 1989; Boreiko \& Betz, 1989; Graf
et al., 1990) to thermalize the low rotational ($J$) lines at a kinetic temperature  
of \gsim$70\,$K (Boreiko et al., 1989).
Under these conditions, one can estimate the opacities of  
the HV gas from the intensity ratio between the $J=2\rightarrow1$ and the 
$J=1\rightarrow0$ lines.  
Fig. \ref{fig:ratio} shows the profiles of the $J=2\rightarrow1$ and the  
$J=1\rightarrow0$ lines of CO and   
the ratio between the lines towards IRc2/I.  
In order to account for the  
different beam size in both lines, the $J=2\rightarrow1$ line has been  
smoothed to the resolution of the $J=1\rightarrow0$ line.

The expected contamination to the $J=1\rightarrow0$ of CO by   
the recombination line H38$\alpha$ is also shown, as a dotted line, in the 
central panel of  
Fig. \ref{fig:ratio}. Since the contribution of the  
recombination lines is negligible for the CO HV gas, we can use  
the CO line ratio over the whole velocity range. 
We are interested only in the relatively compact (\lsim40$''$) emission of the  
HV gas in the IRc2/I region and therefore, the line intensity ratio in  
Fig. \ref{fig:ratio} have been calculated using the main beam brightness  
temperature scale. To obtain the corresponding ratio for extended sources, 
the values in fig. \ref{fig:ratio} should be 
divided by a factor 1.5; i.e. the ratio between the main beam efficiencies 
of the telescope for both lines. 
 
The line ratio shows a remarkably symmetric distribution around 
5$\,$km$\,$s$^{-1}$ (the ambient cloud velocity) suggesting that  
possible contamination of the data from other molecular species emission must be  
negligible. For the ambient velocities of the gas (between 0 and 15$\,$km$\,$s$^{-1}$)  
the emission is extended and the ratio of $\sim1.5$ (1 in the $T_a^*$ scale)  
corresponds to the expected value for extended optically thick emission. As the radial  
velocity increases, the ratio increases up to values around 2.7 for  
velocities of $\mid v_0-v_{\rm LSR}\mid\sim50\,$km$\,$s$^{-1}$.  
For this velocity range our results are   
consistent with those of Snell et al. (1984), once   
the contribution from the extended emission within their larger beam is take into  
account. 
For radial velocities larger than  
$\mid v_0-v_{\rm LSR}\mid\sim50\,$km$\,$s$^{-1}$, the line ratio decreases  
again. The minimum value of 1.5 is close to the values expected for optically thick 
emission. This is present at $\pm55\,$km$\,$s$^{-1}$, just at the  
radial velocity where the bullet features are found (Rodr\'{\i}guez--Franco 
et al., 1999).  
The smaller line ratio is consistent with the increase in the  CO column 
density due to the CO HV bullets. The line ratio rises to its maximum value, 
3.5, for radial velocities near to  
$\pm70\,$km$\,$s$^{-1}$. These data suggest that CO emission for most of  
the velocity range is optically thin, except for the radial velocities where the bullets
are found. Then, except for the CO HV bullets
the CO 
integrated intensity can be translated directly into CO column of  
densities.  
Table \ref{tab:parairc2} gives the physical characteristics for the outflow
and the CO HV bullets. 
The opacity of the bullets is unknown. To derive the properties, 
we have assumed optically thin emission and the ring morphology
observed by Rodr\'{\i}guez--Franco et al. (1999). This gives a lower 
limit to the mass in the CO bullet ring. The total mass in the bullet ring 
represents the {\mbox{\raisebox{-.6ex}{$\stackrel{\scriptstyle >}{\scriptstyle 
\sim}$}}}20\% of the total mass of the outflow.

\begin{table}[bthp] 
\begin{center} 
\begin{tabular}{ccccc}  
\hline 
     & $v_{mt}$\footnotemark[1] & Mass\footnotemark[2] & Momentum & Energy\\ 
     & (km$\,$s$^{-1}$) &($M_{\odot}$)&($M_{\odot}$$\,$km$\,$s$^{-1}$)& (erg)\\ 
\hline 
\hline 
\multicolumn{5}{c}{Wings}\\ 
\hline 
\hline 
blue & $-121$ & 1.10 & 2.3$\times10^{3}$ & 9.86$\times10^{47}$\\ 
red & 142 & 0.96 & 2.5$\times10^{3}$ & 9.88$\times10^{47}$\\ 
\hline 
\hline 
\multicolumn{5}{c}{Bullets}\\ 
\hline 
\hline 
blue & -100 & 0.23 & 23 & 2.30$\times10^{46}$\\ 
red & 80 & 0.43 & 34 & 2.88$\times10^{46}$\\ 
\hline 
\multicolumn{5}{l}{}\\ \cline{1-2} 
\multicolumn{5}{l}{{\footnotesize $^1$ Maximum terminal velocity}}\\
\multicolumn{5}{l}{{\footnotesize $^2$It have been assumed a typical CO abundance of 
 $10^{-4}$}}\\ 
\multicolumn{5}{l}{{\footnotesize  ~and  an excitation temperature of 80$\,$K.}} \\ 
\end{tabular} 
\end{center} 
\caption[]{Physical parameter of the Orion KL molecular outflow and the  
	   associated bullets.} 
\label{tab:parairc2} 
\end{table} 

We have also estimated the physical parameters of the high velocity gas and of the  
bullets associated to the Orion$-$SFBO. We have assumed optically thin emission, 
LTE excitation at a
temperature of 80$\,$K, and a typical CO abundance of 10$^{-4}$. The results  
given in Table \ref{tab:parasur} 
show the characteristics for the molecular outflow and for the bullets. 
The characteristics of these CO HV bullets 
are similar to those found in low mass stars.

\section{A jet driven molecular outflow in the IRc2/I region} 
\label{sec:distribution} 
 
The morphology, the presence of HV bullets, and the high degree of collimation of the
Orion$-$SFBO are clear evidence that this outflow is jet driven, with the jet oriented 
close to
the plane of the sky (see also Rodr\'{\i}guez--Franco et al., 1999). The situation 
for the IRc2/I outflow is less clear and two models (wide opening angle and jet driven
outflow) has been proposed to explain the kinematics and structure of this molecular 
outflow.
Any model proposed to explain the origin of the IRc2/I molecular outflow must account 
for the following observational facts:

\begin{itemize}

\item[\bf a)] The EHV gas does not shows any clear bipolarity around IRc2/I, if any,  
this appears in a position located $20''$ north from this source.

\item[\bf b)] The blue and redshifted HV CO emission show a similar spatial distribution
with an elliptical shape,   
centred near IRc2/I.

\item[\bf c)] There is a systematic trend for the  
size of HV gas to decrease as a function of the radial velocity (see Fig. 
\ref{fig:IRc2-vel}).

\item[\bf d)] The presence of CO HV bullets distributed in a thin elliptical ring-like 
structure around the EHV gas (see Fig. \ref{fig:bipolar}e and f) surrounding the EHV gas.

\end{itemize}

We will now discuss how the proposed models account for these observational 
results.

\subsection{Wide open angle conical outflow powered by source I}

The model used to explain the IRc2/I outflow at moderate velocities is 
a wide open angle
($\sim130^{\rm o}$) conical outflow oriented in the 
southeast$\leftrightarrow$northwest direction and
powered by source I (Chernin \& Wright, 1996). 
This model was based on the weak bipolarity of the MHV in the Orion IRc2 outflow 
reported by Erickson et al. (1982).
Observations with higher angular resolution of the MHV CO emission seem to support 
this type of model (Chernin \& Wright, 1996). Recent VLA observations of the SiO maser
distribution can also be explained by this kind of model.
However, the kinematics and the morphology of the low velocity H$_2$O
maser emission cannot be explained by the SiO outflow.
Two alternatives have been suggested: an additional outflow powered by the 
same source and expelled perpendicular to that producing the SiO
masers (Greenhill et al., 1998), and a flared outflow (Doeleman et al., 1999).
Furthermore, from the morphology of the HV gas, Wilson et al. (1986) pointed out  
that if the HV CO emission were produced by a wide open angle bipolar outflow driven by 
IRc2/I, 
it would require a complex channeling of the outflowing gas to account for  
the morphology of the MHV CO emission. This situation is even more extreme when  
the morphology of the EHV gas presented in this paper is considered
(see also Rodr\'{\i}guez--Franco et al., 1999). If the outflow is wide opening angle 
one could use the morphology of the EHV gas to locate the powering source as in 
the case of low mass stars (see Chernin \& Wright, 1996). 
If so, the source(s) driving the outflow should be $20''$ north of IRc2. 
If the exciting source of the EHV is located $20''$ north of IRc2, then 
IRc2 would not be the cause of the CO outflow (see Figs. \ref{fig:bipolar}c and d).  
Similar arguments would also apply to other wide opening angle wind models like those of
Li \& Shu (1996).

\subsection{Multiple molecular outflows scenario}

One alternative to explain the CO morphologies, 
would be several molecular outflows in the region. A wide open angle bipolar
outflow powered by source I with moderate velocities in the CO emission and a compact
highly collimated and very fast outflow powered  by an undetected source located 
approximately 
$20''$ north of IRc2 (EHV outflow). However, such a multiple outflow scenario would 
require an additional outflow to explain the low velocity H$_2$O 
maser emission. This would be one of the highest concentration of outflows found 
in a star forming region.

Although the multiple outflow scenario is possible, 
it seems certain that 
all (the HV CO emission,
the low and high velocity H$_2$O 
masers in the IRc2 region, the SiO masers, and the HV CO molecular bullets)
the observational features of the IRc2/I outflow could be
caused by a molecular outflow driven by a variable precessing jet oriented 
along the line of sight and powered by 
source I (Johnston et al., 1992; Rodr\'{\i}guez--Franco et al., 1999).

\subsection{A molecular outflow driven by a wandering jet} 
\label{sec:jet}

Rodr\'{\i}guez--Franco et al. (1999) have presented a number of arguments in 
favor of the possibility that the IRc2 outflow is driven by a wandering jet
oriented along the line of sight.
To strengthen the arguments for this model, 
we will analyze in detail the 
size--terminal velocity dependence found in the previous section, and the mass
distribution of the HV gas as a function of the radial velocity. 

\subsubsection{The size--velocity relation}
 
\begin{figure*} 
   \vspace{8.5cm}
   \includegraphics{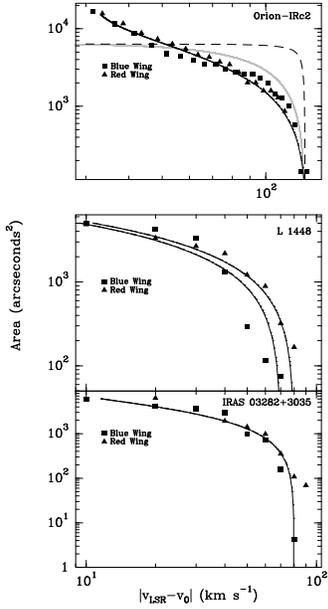} 
   \caption{{\bf a)}, {\bf b)}, and {\bf c)} The distribution of the area enclosed 
            by equal projected terminal  
	    velocity as a function of the projected terminal velocity for the  
	IRc2, L$\,$1448, and I$\,$3282  bipolar outflows, respectively. 
	For the three panels, 
	the filled squares and filled triangles represents, respectively, 
	the observational data for blue and red line wings. The fits are determined from 
        the model of a jet driven molecular outflow with the
	parameters in Table 3. This is represented by a solid curve in 
	each panel. 
	In the Orion IRc2 panel we have also represented the expected distribution
	for an spherical 
	expansion model and an expanding ellipsoid model
	(dashed line). 
       } 
     \label{fig:modeloa} 
  \end{figure*}

In Fig. \ref{fig:modeloa}a we present the dependence of the area enclosed level 
at a given 
terminal velocity as a function of the terminal velocity (see Figs. \ref{fig:bipolar}c
and d).
As already noted, the blue (filled squares) and the 
redshifted (filled triangles) 
HV gas show very similar distributions.
This similarity  
could be due to a constant spherical or elliptical expansion at constant velocity, 
as that proposed for the H$_2$O masers in Orion and in other massive star 
forming regions 
(Genzel \& Downes, 1983; Greenhill et al., 1998). However, the expected
area$-$velocity dependence for this kind of expansion (see the dashed 
and dotted lines in Fig. \ref{fig:modeloa}a) is inconsistent with the CO data, 
indicating that the isotropic low velocity outflow modelled for the H$_2$O
masers do not appear in the bulk of the HV gas. Then the H$_2$O masers only 
represents a small fraction of 
the outflowing gas. We therefore exclude this possibility for the CO emission.

We now consider that the observed size$-$velocity distribution 
is produced by a jet oriented along the line of sight. 
First, we compare the size$-$velocity  
distribution observed for the IRc2/I outflow with other low mass outflows which 
are known  
to be driven by jets. Unfortunately, The Orion--SFBO outflow cannot be used 
because perpendicular to the jet it is only slightly resolved
by our beam. 

Good examples of jet driven outflows are L$\,$1448 and 
IRAS$\,03282+3035$ (hereafter I$\,$3282)  
(see e.g. Bachiller et al., 1990, 1991). 
These two outflows are driven by low mass stars and their jets are  
oriented at a small angle ($\leq 45^{{\rm o}}$) relative to the plane of the sky.  
To compare the area$-$velocity distribution measured in these outflows with 
that expected when the jet is aligned along the line of sight,
we have to rotate 
the outflow axis to point towards the observer. To do this, we have assumed 
that the outflows have a cylindrical geometry. The radial velocities have not
been corrected for projection effects since this is a constant factor for all
velocities.
The results for L$\,1448$ and I$\,$3282 are shown in Figs. \ref{fig:modeloa}b 
and c respectively. Remarkably, the area$-$velocity dependence 
derived for both jet driven molecular outflows
are very similar to that found for the IRc2/I outflow. 
These results strongly support the idea that the IRc2/I outflow is also driven 
by a jet oriented along
the line of sight.

To explain the area$-$velocity dependence found for these outflows,
we have considered a very simple model which mimics the kinematics of a 
bipolar jet. In this simple model, the ejected material 
is very fast and well collimated around the symmetry axis. 
Away from the jet axis, the material surrounding the jet is entrained
generating the more extended low velocity outflow with lower 
terminal velocities. 
We have modeled this kinematics by using a simple velocity law  given by

\begin{equation} 
{\buildrel\rightarrow\over v}(x,y,z)=v_{jet}\exp^{
\left(-\frac{x^2+z^2}{2p^2}\right)}{\buildrel\rightarrow\over
\jmath}
\label{eq:ley0} 
\end{equation}

\noindent
where $x$, $y$ and $z$ are the spatial coordinates, $p$ is the 
collimation parameter of the outflow (i.e. the radius of the jet),
$v_{jet}$ is the velocity of the 
molecular jet and ${\buildrel\rightarrow\over \jmath}$ is the direction in which the
material is moving.   When the material is within the jet ($x^2+z^2\leq2p$) 
${\buildrel\rightarrow\over \jmath}$ is along the jet direction, 
while it is in the radial direction outside the jet. The bipolar morphology is 
taken into account by supposing 
that the HV gas is restricted to a biconical geometry. 

The model combines two kind of parameters: 
the intrinsic outflow parameters such as 
$p$ and $v_{jet}$,
and the geometry (i.e. the cone parameters). 
The collimation parameter is derived by fitting the observations, and 
$v_{jet}$, which cannot be determined since the orientation of the jet along 
the line of sight is unknown, has been considered the 
terminal velocity measured for the data.
If the symmetry axis 
of the cone is along the line of sight,  the
semi-axis of the ellipses on the plane of the sky ($a$ and $c$) are directly 
measured from the size corresponding to the minimum terminal 
velocity contour (see Fig. \ref{fig:bipolar}c and d). 
The length of the cone along the line of sight,  
$b$, is a free parameter, determined from the model. 
Under the assumptions discussed
above, the model contains only two free parameters: 
the collimation parameter,
and the size of the molecular outflow along the line of sight.

The results of this simple model for the best fit to the data for the 
three outflows are shown in Figs. \ref{fig:modeloa}a,
b, and c as solid lines, and the derived parameters are given in table 
\ref{tab:parametros}.
The results are very sensitive to small changes of the
collimation parameter, but very insensitive to the size of the outflow
along the line of sight. Changes by $20\%$ in the collimation parameter 
greatly worsen the fit to the data. The data, however, can be fit with any length 
of an outflow
larger than the value of the minor semiaxis of the ellipsoid. 
Our results indicate a similar collimation parameter for the two 
outflows driven by low mass stars, but it is a factor of 2 larger 
for the IRc2/I outflow. This difference can be related either to the
fact that the low 
mass star outflows are much younger than that of the IRc2/I outflow, or to 
different types of stars driving the outflows. 
We conclude that the overall kinematics and the morphology of the EHV gas observed in 
the IRc2/I outflow is consistent with a jet driven molecular outflow oriented 
along the line of sight with a jet radius of 0.06$\,$pc.  

\begin{table}[bthp] 
\begin{center} 
\begin{tabular}{c|c|c|c|cc|c|c} 
\hline 
Outflow & $a$ & $b$ & $c$ &  \multicolumn{2}{c|}{$p$} &$v_{min}$ & $v_{jet}$ \\ 
    &$''$&$''$&$''$&$''$&pc&km$\,$s$^{-1}$&km$\,$s$^{-1}$ \\ \hline
IRc2    & 50  & 70 & 100 & 25 & 0.06 &  31       & 101        \\
L$\,$1448& 40 & 40  & 150 & 20 & 0.03 &  0       &75{\footnotesize $^b$}-85{\footnotesize $^r$}
  \\
I$\,$3282& 44 & 44  & 235 &  20 & 0.03 &  5       & 75  \\ \hline
\multicolumn{8}{l}{}\\ \cline{1-3} 
\multicolumn{8}{l}{{\footnotesize $^b$For the blue wing}}\\ 
\multicolumn{8}{l}{{\footnotesize $^r$For the red wing}}\\ 
\end{tabular} 
\end{center} 
\caption{Parameters used to fit the velocity law.} 
\label{tab:parametros}  
\end{table}

\subsection{The mass distribution of the gas} 
\label{sec:masa}

We have shown that a jet driven molecular outflow can explain the
terminal velocities and the spatial distribution of the HV gas. The next
question is, how does this model explain the mass
distribution of the HV gas in the outflow?.

\begin{figure*} 
   \vspace{6cm}
   \includegraphics{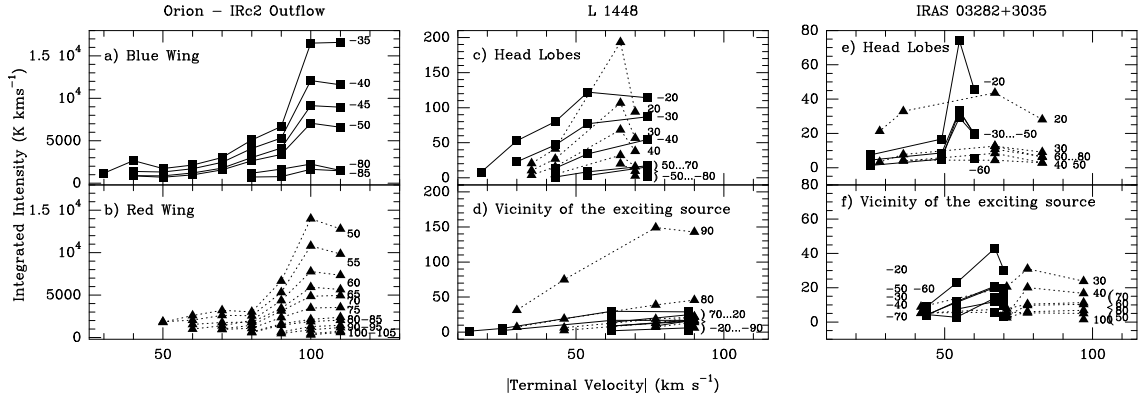} 
   \caption{Panels a) and b): dependence of the mass distribution in velocity intervals 
	    of 5$\,$km$\,$s$^{-1}$ (CO  
	    integrated intensity in velocity intervals of 5$\,$km$\,$s$^{-1}$ wide) 
	    with the terminal velocity (distance to the outflow axis)
	    for the IRc2 bipolar outflow. Panel a) corresponds 
	    to the blue wing while panel b) to the red wing. In both 
	    lobes one can distinguish the region associated to the shock  
	    (bow shock) and the region associated to the high-velocity jet.  
	    Numbers in the right side of every curve are the value in  
	    $\,$km$\,$s$^{-1}$ of the superior extreme of considered velocity  
	    interval.
	    \newline
	Panels c), d), e), and f): dependence of the mass distribution in velocity 
	intervals of 10$\,$km$\,$s$^{-1}$ 
	(CO integrated intensity by velocity interval) with the terminal 
	velocity for the blue (squares) and red (triangles) wings of the 
	L$\,$1448 (panels c) and d)) and I$\,$3282 (panels e) and f))outflows. 
	Panels c) and e) 
	correspond to the region which is far from the exciting source. 
	It is in this region where probably a shock between the ambient gas 
	and the ejected  
	material is produced (bow shock). Panels d) and f) correspond to the 
	most close region to the exciting source. Is 
	in this region where higher terminal velocities are observed. 
	Each curve has been noted with the most negative (blue wing) and most 
	positive (red wing)
	value in $\,$km$\,$s$^{-1}$ of the velocity interval considered. 
       } 
     \label{fig:masa-flujos} 
  \end{figure*}

In Figure \ref{fig:masa-flujos}a and b we show  masses derived from the CO line
intensities
integrated in 5$\,$km$\,$s$^{-1}$ intervals as a function of velocity. 
This figure shows that the bulk of the mass at moderate radial velocities is also 
found at the locations where the gas shows the largest terminal velocity. 
In the proposed jet driven model, this would correspond to the material 
located at small projected distance from the outflow axis (i.e. in the jet 
direction). 

At first glance, these results are surprising since one expects to find 
only the highest velocity gas in the jet direction. However, there are two possibilities 
which could explain the derived mass distribution in the IRc2/I outflow. 
In the first, one assumes 
that the vicinity of the exciting source large quantities of gas 
moves with all velocities as a result of the dragging of the ambient 
material by the action of the jet. In the second, one assumes that just the 
opposite is true, i.e. the masses at moderate velocities are located far from 
the exciting 
source, only in the jet heads, where jet impacts on the ambient medium.
In this case, the HV gas will appear with all velocities only in the direction
of the jet. 
Unfortunately, the orientation of the IRc2/I outflow, 
along the line of sight, prevent us from determining which of the two proposed 
scenarios account for the data. Again, we will 
compare the results for the IRc2/I outflow with those
obtained from the jet driven outflows L$\,$1448 and I$\,$3282 as described in 
section 6.3.1.

We have estimated the expected mass distribution of the L$\,1448$ and 
I$\,3282$ outflows if the jets were oriented along the line of sight. For
these two outflows it is possible to separate the 
contribution to the total mass of the outflow for two different regions
in the outflow. The head lobes, and the exciting source.
Figs. \ref{fig:masa-flujos}c to e show the
mass distribution as a function of the radial velocity in these two regions 
for the L$\,1448$ and the I$\,3282$ 
outflows in velocity intervals of 10$\,$km$\,$s$^{-1}$.

Using the results of L$\,$1448 and I$\,$3282, the bulk of the 
mass for all terminal velocities would arise from the head lobes close to 
the jet axis where one also observes the largest terminal velocities. Only a 
small fraction of the mass is found close to the exciting source.
These findings are consistent with the observations of the IRc2/I outflow and
implies that the largest fraction of the mass of the outflowing gas  
at low velocities is preferentially concentrated in the regions at the head 
lobes, and close to the jet axis.
The similarity between the results obtained for the bipolar outflows 
L$\,1448$ and I$\,3282$ and those found in the IRc2/I outflow suggest 
that the three outflows have a very similar mass distribution. 
This would indicate that the massive outflow driven by source I is produced 
by a jet and
that a large fraction of the outflowing gas at low or moderate velocities is 
located at the end of jets in the head lobes, 
just in front and behind of the exciting source.

\section{The interaction between the jet and the ambient material} 
\label{sec:discusion}

\begin{figure*} 
   \vspace{5cm}
   \includegraphics{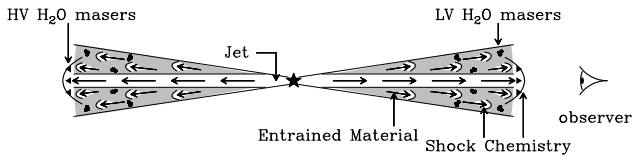} 
   \caption{A sketch of the proposed model for the IRc2/I outflow.
   The outflow direction has been rotated by $90^{\rm o}$ with respect the line of sight.
   The filled star
   represents the powering source, triangles and dots the high and low velocity 
   H$_2$O masers respectively and the arrows the direction of the wandering
   jet in its different episodes. In the head of the jets, bow shocks are represented. 
   It is in these regions where shock chemistry can dominate. 
           } 
     \label{fig:modelo} 
  \end{figure*}

The precessing jet scenario proposed for the molecular outflow in Orion has 
important consequences in order to explain the different phenomena produced
by the interaction of the outflow and the ambient gas like the H$_2$ 
vibrationally excited emission (H$_2^*$), the shock-chemistry found in 
this region, and the location and the origin of the low velocity and 
high velocity H$_2$O masers. Figure \ref{fig:modelo} shows a sketch 
of the proposed model showing the regions 
where the different emissions could arise. According to this model 
(Rodr\'{\i}guez--Franco et al., 1999), 
a precessing jet with very high 
velocities (larger than 100$\,$km$\,$s$^{-1}$) is aligned close to the line of 
sight. The HV jet interacts with the 
surrounding material sweeping a considerable amount of gas and dust. The 
impact of the precessing jet on the ambient molecular gas produces a number of
bow shocks 
in the head of the two lobes. Within the most recent bow shocks one expects to 
observe H$_2^*$ emission and bullets 
and the low and high velocity H$_2$O masers as discussed in detailed by 
Rodr\'{\i}guez--Franco et al. (1999). A precessing jet would explain the 
distribution of the HV bullets and the large number of H$_2$O masers at low
radial velocities. We now will discuss how the proposed 
model can also explains the overall properties of the H$_2^*$ emission and the 
``plateau'' emission.

\subsection{The H$_2^*$ emission}

\begin{figure*} 
   \vspace{3.5cm}
   \includegraphics{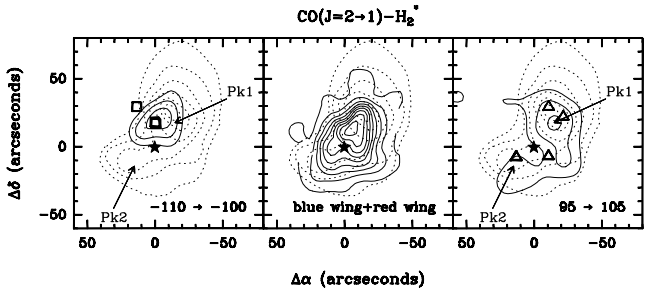} 
   \caption{Comparison between the blue wing $J=2\rightarrow1$ CO emission 
	and vibrationally excited molecular hydrogen  
	emission (Beckwith et al., 1978) For each panel, CO emission is 
	represented by continuous contours while H$_2^*$ is represented  
	by dashed contours. Left and right panels show the positions noted 
	as Pk1 and Pk2 by Beckwith et al. (1978). In these panels, emission 
	corresponding to the largest velocities are represented 
	(left, blue and right, red). Also positions of HV  
	water masers are noted (squares represent negative 
	velocities or blueshifted, while triangles represent positive  
	velocities or redshifted). Velocity intervals of integration  
	appear in the bottom right corner. First level is  
	1$\,$K$\,$km$\,$s$^{-1}$ and  
	interval between levels is 0.5$\,$K$\,$km$\,$s$^{-1}$. The CO 
	emission in central panel represents the addition of the emissions 
	between $-100$ and $-30\,$km$\,$s$^{-1}$ plus  
	45 and 95$\,$km$\,$s$^{-1}$. First level is 15$\,$K$\,$km$\,$s$^{-1}$  
	and interval between levels is 15$\,$K$\,$km$\,$s$^{-1}$. 
	For all the panels the filled star represents the IRc2 position.  
       } 
     \label{fig:co-h2} 
  \end{figure*}

The most straightforward evidence of the interaction between the high 
velocity jets and the ambient medium comes from H$_2^*$ (see e.g., 
Garden et al., 1986; Doyon \& Nadeau, 1988). In our model, the H$_2^*$ emission 
should mostly appear at the head lobes (see Fig. \ref{fig:modelo}). 
The H$_2^*$ emission is expected to be located just in the intermediate layer 
between the jet and the HV CO emission (see Rodr\'{\i}guez--Franco et al., 1999), and 
one would, therefore, expect similar extent for both emissions. This is illustrated in the central panel of 
Fig. \ref{fig:co-h2} where we present the comparison between the spatial 
distribution of the CO emission in the bow shock and that of the H$_2^*$. In the proposed 
geometry, the 
H$_2^*$ emission should be affected by the extinction produced by the HV 
blueshifted gas and dust located between the observer and the redshifted 
H$_2^*$ layer. Indeed, 
though the overall H$_2^*$ and CO bow shock emissions are very similar, there 
are also important differences. 
As shown in Fig. \ref{fig:co-h2}, the 
most intense CO bow shock emission is located between IRc2 and the H$_2^*$ Peak 1 
(Beckwith et al., 1978), just in the region where the H$_2^*$ emission shows a 
minimum. This difference agrees with the idea of a large accumulation of 
material near the jet heads, close to the direction where the jet impinges.

From our CO data, the largest column of density in the blueshifted bow shock is found
$12''$ north of IRc2/I. At this position we derive, from the 
HV CO data, an H$_2$ column of density of $4.5\times10^{21}\,$cm$^{-2}$  
which corresponds to an extinction at 2.1$\,\mu$m of $0.5\,$mag, in 
good agreement with the 
extinction of 0.6$\,$mag derived  by Geballe et al. (1986).
Additional evidence in favor of the proposed geometry comes from the 
variation of the extinction as a function of the velocity derived from the 
H$_2^*$ emission. Scoville et al. (1982), and Geballe et al. (1986), have 
found that the extinction in the H$_2^*$ line wings is larger than at the 
line center by $\sim1\,$mag.
If the ejection is jet-driven, a large quantity of 
ambient material is accumulated in the head of the lobes near the outflow
axis where one observes the largest radial velocities.
Therefore, if the jet axis is 
oriented along the line of sight, the highest velocities should be the most 
affected by the extinction produced by the gas and dust pushed by the outflow. 
In a similar way, the proposed geometry can naturally explain the asymmetry 
observed in 
the H$_2^*$ line emission in which the blueshifted emission is less extincted 
(between 0.1 and $1.5\,$mag) than the redshifted emission.

With the proposed scenario, one can estimate the ratio between the 
H$_2^*$ and the CO HV material in the bow shock. For the position of Pk1, the 
vibrationally excited hydrogen column density is 
$\sim6\times10^{17}\,$cm$^{-2}$ (Brand et al., 1988), and the CO column 
density obtained for both outflow wings is $\sim1.8\times10^{17}\,$cm$^{-2}$. 
Then, the CO/H$_2^*$ ratio would be $\sim0.2-0.3$. This indicates that, 
as expected, only a very small fraction of the shocked gas is hot 
enough to be detected in the H$_2^*$ lines. Even, after the correction for 
extinction, the thickness of the H$_2^*$ layer must be at least two orders of 
magnitude thinner than the colder shocked CO layer.

\subsection{The ``plateau'' emission}

The ``plateau'' component is the source of broad line wings in a number of 
molecules which are believed to be produced by shock chemistry. 

Based in observations of molecules like SO, SO$_2$, SiO and HCO$^+$, which are good 
tracers of the low-velocity outflow, Plambeck et al. (1982) and Wright et al. 
(1995) have suggested that this low velocity emission arise from a ring or 
``doughnut'' of gas expanding outward from IRc2/I. The origin of these molecules 
is closely relate with shocks (see Mart\'{\i}n--Pintado et al., 1992),
and they are produced by the interaction of the outflowing gas with the 
dense ambient clouds (Bachiller \& P\'erez--Guti\'errez, 1997).

In the proposed model, the bulk of shocked gas occurs in the bow shocks 
produced in the heads of the two lobes, in the direction of the line of 
sight. As in the low mass outflows (Bachiller \& P\'erez--Guti\'errez, 1997), 
the bow shocks will favor the observation of molecules 
characteristics of a shock chemistry like HCO$^+$, SO, SO$_2$ or SiO in the region of 
high density. This might explain why this emission has low velocity and is 
observed just around the outflow axis. 
A similar situation have been observed in L$\,$1157, a low mass star driven 
outflow, whose axis is almost in the plane of the sky (Bachiller \&
P\'erez--Guti\'errez, 1997). In this outflow one can observe that the abundance of 
molecules like SiO, CH$_3$OH, H$_2$CO, HCN, CN, SO and SO$_2$ is enhanced by 
at least an order of magnitude in the shocked region at the head of 
both lobes, with low abundance in the vicinity of the exciting source.

\subsection{Models of jet-driven molecular outflows}

Most of the mass observed in molecular outflows is made by ambient material 
entrained by a ``primary wind'' from the
central source. Two basic processes of entrainment has been considered to explain the
bulk of the mass in the bipolar outflows (see e.g. Cabrit, 1995).

\begin{itemize}

\item[a)] Viscous mixing layers at the steady-state,
produced via Kelvin--Helmholtz (KH) instabilities at 
the interface between the outflow and the ambient cloud (Stahler, 1994). 

\item[b)] Prompt entrainment, produced at the end 
of the jet head in a curved bow shock that accelerates and sweeps the ambient 
gas creating a dense cover with a low density cocoon surrounding the jet 
(Raga \& Cabrit, 1993).
\end{itemize}

Studies of the CO line profiles in several molecular outflows indicates that prompt
entrainment at the jet head is the main mechanism for molecular entrainment 
(see e.g. Chernin et al., 1994).
Furthermore, precessing jets have been also invoked to explain the large opening angles of
bipolar outflows (Chernin \& Masson, 1995). However, it has been argued that the 
precessing
angles are small and the propagation of large bowshock  seems to be the dominant 
mechanism in the formation of bipolar outflows for low mass stars (Gueth et al., 1996).

The proposed scenario (a jet driven molecular outflow) for the IRc2/I outflow confirms
that the main mechanism for entrainment in this outflow is also prompt entrainment at 
the jet
heads, in agreement with the Raga \& Cabrit (1993) model. However, the presence of HV 
bullets (see Figs. \ref{fig:bipolar}e and f) are best explained (see
Rodr\'{\i}guez--Franco et al., 1999) by the model of a 
precessing jet (Raga \& Biro, 1993). Furthermore, the area-velocity relation is well 
explained in both, low mass
star molecular outflows (L$\,$1448, I$\,$3282), and in the massive star IRc2/I outflow by a
radial expansion from the exciting source similar to that found in the L$\,$1157 outflow
(Gueth et al., 1996).
As illustrated in the sketch in Fig. \ref{fig:modelo} this would be easily explained in 
the 
framework of the precessing jet with prompt entrainment since the shocked material
will be always moving in the direction of the jet, i.e., just in the radial direction
from the exciting source.

Another important result from our data is the large transverse velocities measured
from the CO data which can be up to $20-30$\% of the jet velocity. 
The large transverse velocities in jet driven molecular outflows will also explain the
shell-like outflows as more evolved objects. The typical time scale for a young jet-like
molecular outflow to evolve to a shell-like molecular outflow with a cavity of 
$\sim3\,$pc will be of $\sim10^5\,$years for the typical transverse velocity measured in the
IRc2/I outflow. This is in agreement with the ages found in shell like
outflows powered by intermediate mass stars (NGC$\,$7023, Fuente et al., 1998) and low mass stars
(L$\,$1551-IRS$\,$5, Plambeck \& Snell, 1995).

\section{Conclusions}

We have mapped the Orion region in the $J=2\rightarrow1$ line of CO with the 30$\,$m
telescope. From these maps we have detected high velocity gas in two regions: the
IRc2/I outflow and the Orion--S outflow. 
The main results for the Orion--S outflow can be summarized as follows,

\begin{itemize}  
 
\item The bipolar molecular outflow in the Orion-S region presented in this paper is very fast
($\sim110\,$km$\,$s$^{-1}$) and compact (\lsim$0.16\,$pc). It is highly collimated and shows the
presence of HV velocity bullets. It is perpendicular to the monopolar low velocity 
($<30\,$km$\,$s$^{-1}$) outflow known in this region (Schmid$-$Burgk et al., 1990).

\item The location of the possible exciting source is estimated, from the kinematics 
of the
high velocity gas, to be $20''$ north from the position of FIR$\,4$. At 
this position no
continuum source in the cm or mm wavelength range has been detected.

\item The morphology of this bipolar outflow suggests a very young (dynamical age of
$\sim10^3\,$years) jet driven molecular outflow similar to those powered by low mass
stars.

\end{itemize}

For the IRc2/I outflow the main results are:

\begin{itemize}

\item While the HV gas with moderate velocities 
($v_{LSR}\leq \mid 55\mid\,$km$\,$s$^{-1}$) 
is centred on 
IRc2/I with a very  weak bipolarity around IRc2/I, the morphology of the blue and 
redshifted HV gas for the most extreme velocities 
$\mid v_{LSR}-9\mid$\lsim80$\,$km$\,$s$^{-1}$ (EHV) peaks $20''$ northwest from IRc2/I
and $10''$ south of IRc9. The EHV gas does not show any clear bipolarity around IRc2/I. 
The only possible bipolarity in the east$\leftrightarrow$west direction is
found $20''$ north of IRc2. The blue and redshifted HV CO emission show a similar spatial
distribution with an elliptical-like shape centred in the vicinity of IRc2/I, and a
systematic trend of the size of the HV gas to decrease as a function of the velocity.

\item The morphology and kinematics of the HV CO emission cannot be accounted by the most
 accepted model:
the wide opening angle outflow. We discuss other alternatives such as multiple outflows
and a precessing jet driven molecular outflow oriented along the line of sight. 

\item We have compared the size-velocity dependence and the mass 
distribution in the Orion IRc2/I outflow
with those derived from the jet driven molecular outflows powered by low mass stars 
(L$\,1448$ and I$\,3282$) when these are projected to have the jet oriented along the line
of sight. We find very good agreement between the jet driven molecular outflow in
low mass stars with that of the Orion IRc2/I outflow indicating that this outflow can be 
jet driven.

\item The size-velocity dependence found for the outflows is fit 
using a simple velocity law which consider the
presence of a highly collimated jet and entrained material.
The velocity decreases exponentially from the jet and it is in the
radial direction for the entrained material outside the jet. We derive similar
collimation parameters for the L$\,1448$ and the I$\,3282$ outflows of 0.03$\,$pc, a 
factor of two larger for
the Orion--IRc2/I outflow. This difference might be an age effect, or due to the
different type of exciting stars.

\item From a comparison of the mass distribution as a function of velocity, we conclude
that the bulk of the HV gas in the Orion IRc2/I outflow is produced by prompt entrainment
at the head of the jet.

\item The morphology and kinematics of the shock tracers, H$_2^*$, H$_2$O masers,
H$_2$ bullets, the ``plateau emission'', and the radial direction of the entrained HV gas
in the IRc2/I outflow
is qualitatively explained within the scenario of a molecular outflow driven by 
a precessing jet
oriented along the line of sight. The large transverse velocity found in this outflow can
explain the shell-type outflows as the final evolution of the younger jet driven
outflows. 

\end{itemize}

  \acknowledgements 
   This work was partially supported by the Spanish CAICYT under grant  
   number PB93-0048. 
  
%

\newpage

  \begin{figure*} 
   \vspace{9.5cm}
   \includegraphics{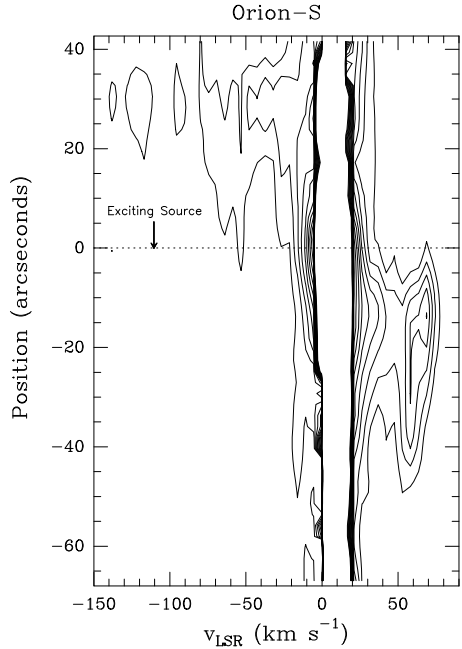} 
   \caption{Velocity$-$position strip in the direction defined by the symmetry 
	major axis of the Orion$-$SFBO. Vertical scale  
	represents positions relatives to geometric centre of the outflow over 
	the mentioned axis. Dashed horizontal lines are traced in the  
	positions of the exciting source and the 1.3$\,$mm continuous  
	source. The lowest level corresponds to an intensity of 0.13$\,$K; 
	the distance between levels is 0.2$\,$K. 
       } 
     \label{fig:corte-sur} 
  \end{figure*}

\end{document}